# Electronic Visualisation in Chemistry: From Alchemy to Art

Karl Harrison
University of Oxford
Department of Chemistry
United Kingdom
http://www.chem.ox.ac.uk/it/karlharrison.html
*karl.harrison@chem.ox.ac.uk*

Jonathan P. Bowen
London South Bank University
Department of Informatics
United Kingdom
http://www.jpbowen.com
*jonathan.bowen@lsbu.ac.uk*

Alice M. Bowen
University of Oxford
Department of Chemistry
United Kingdom
http://oxford.academia.edu/AliceBowen
*alice.bowen@chem.ox.ac.uk*

**Chemists now routinely use software as part of their work. For example, virtual chemistry allows chemical reactions to be simulated. In particular, a selection of software is available for the visualisation of complex 3-dimensional molecular structures. Many of these are very beautiful in their own right. As well as being included as illustrations in academic papers, such visualisations are often used on the covers of chemistry journals as artistically decorative and attractive motifs. Chemical images have also been used as the basis of artworks in exhibitions. This paper explores the development of the relationship of chemistry, art, and IT. It covers some of the increasingly sophisticated software used to generate these projections (e.g., UCSF Chimera) and their progressive use as a visual art form.**

*Electronic visualisation. Virtual chemistry. Molecular structure. 3D projection. Digital art.*

## BACKGROUND

From the very beginnings of the understanding of chemical structure, chemists have had to use visualisations to represent their view of their science. Chemistry is the creation and study and properties molecular and chemical structures built from the bonding of atoms, these atoms are composed of infinitively small atomic nuclei and clouds of electrons in probability fields and where bonds are formed from the overlap of these electronic clouds of quantum distribution. None of this can be seen experimentally and only inferred from the perturbation and diffraction of x-ray or neutron beams. As a result chemists have devised schematics to illustrate the chemical properties and reactions of the molecules they study. These schematics of course are used in the research publications and textbooks to explain and discus chemistry. Originally, chemical schematics could be produced using typewriters and simple offset printer plates (Figure 1).

Quite rapidly, these simple representations of chemical structure became unsatisfactory as the knowledge of chemical structure evolved. Illustrators became employed either at publishers or even embedded into science departments to produce the schematic illustrations for the science community (Figure 2).

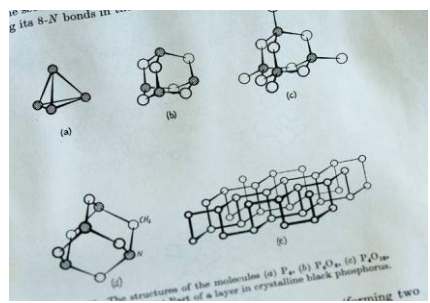

**Figure 2:** *Early illustration of structures (Wells 1956)*

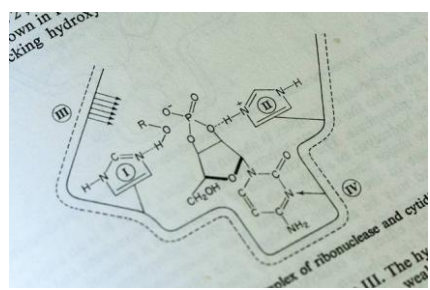

**Figure 3:** *Representation of chemical to protein interactions (Goodwin 1964)*

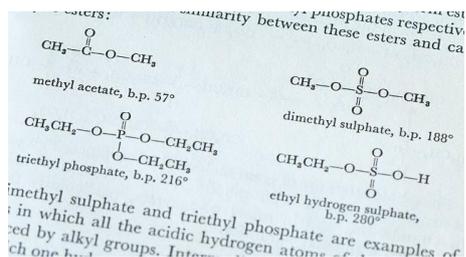

**Figure 1:** *Typeset chemical structures (Grundon 1962)*

One of most famous illustrators in this field was Irving Geis who worked with publishers to illustrate chemistry and biochemistry structures. In 1961, he was commissioned by *Scientific American* to illustrate John Kendrew's article of the first protein structure, Myoglobin, followed by illustrations for the article by David Phillips (1966) on the first enzyme structure, Lysozyme (Figure 4).

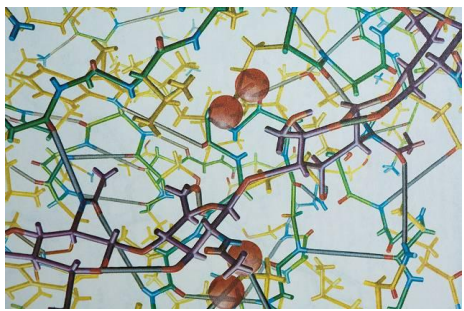

*Figure 4: Painting of Lysozyme (Philips 1966)*

In 1976 Irving Geis co-authored a beautiful textbook with Richard Dickerson entitled *Chemistry, Matter, and the Universe* (Dickerson & Geis, 1976). See Figures 5 and 6.

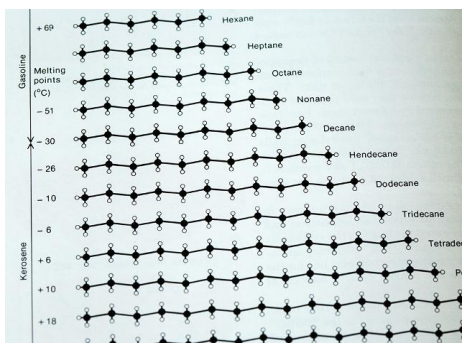

*Figure 5: Organic alkane chains (Dickerson & Geis 1976)*

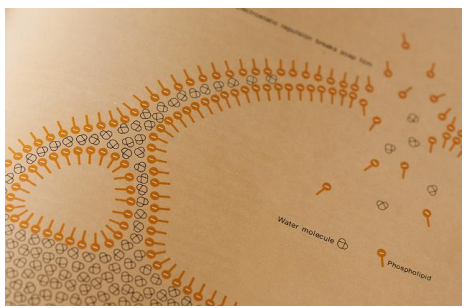

*Figure 6: The chemistry of soap bubbles (Dickerson & Geis 1976)*

However the rapid rate of chemistry publications and growth of scientific knowledge resulted in the chemists needing to produce themselves camera-ready illustrations for their research papers and books.

So from the late 1970s to mid 1980s, many science authors became expert using Rotring pens™, plastic templates stencils and Letraset™. Thus many hours would be spent by untrained chemists but now illustrators trying to produce camera-ready illustrations using these tools (Figure 7), with a constant worry of smudging the ink or tearing a piece of Letraset™ transfer that could force them to start over again.

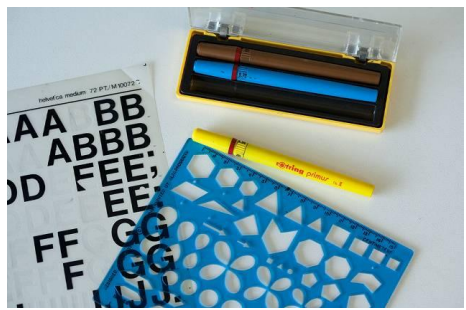

*Figure 7. Chemistry illustration tools*

1984 was truly "1984" for chemists; Apple introduced the Apple Mac and this was revolutionary. The original Apple Mac came with a grey scale-screen and a functioning mouse to allow drawing and was shipped with MacDraw and MacPaint. But importantly the first Apple Mac came with the ability to print Adobe postscript to the very first Apple Laser-printer. This was the game changing innovation; for the first time, a computer could be used to produce camera-ready illustrations for print and publication

In 1986, Cambridgesoft was founded and released ChemDraw for Mac. This rapidly prompted chemists around the world to buy Apple Mac computers and to this day desktop computing in chemistry departments is still dominated by Apple Mac computers. In 1994, Cambridgesoft released ChemDraw 3.1 for Windows, and this allowed the then new Microsoft Windows computers using MS Windows 3.1, to have a chemistry drawing solution for their computer platform. Cambridgesoft's ChemDraw is still very much the publishing standard and many journals will only accept chemical illustrations produced in this propriety file format. Complex chemical schematics and illustrations are rapidly produced in ChemDraw and it has inbuilt chemical intelligence knowing structural rules, error highlighting and auto-drawing/correction functions from formulae, structural names or even poorly drawn figures (Figure 8).

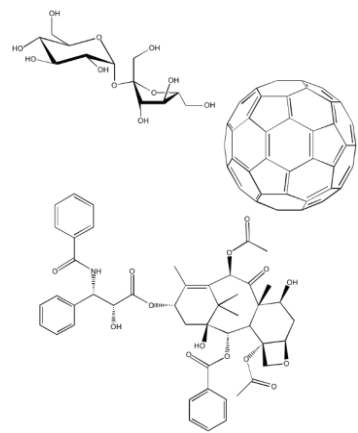

*Figure 8: Chemical drawings of sucrose, C60 and Taxol*

Chemistry though is not just study of two-dimensional line drawings but very much a three-dimensional subject. Traditionally the 3D aspects were covered by tactile 3D models, originally made in wood, then metals and then with plastic atoms and bonds (Figure 9).

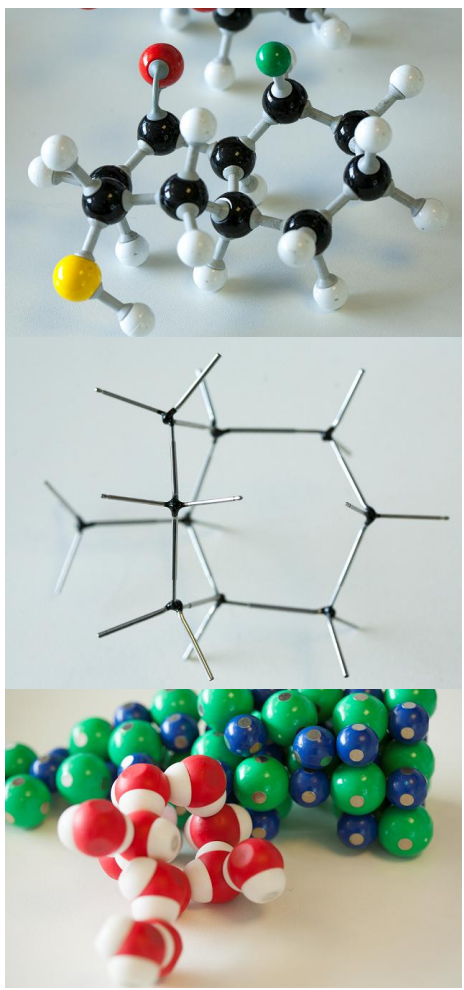

*Figure 9: Plastic and metal models of chemical structures*

**ELECTRONIC VISUALISATION**

Chemistry software was quickly produced as soon as desktops computers became popular to provide chemists the ability to visualize their structures in 3D on their computer screens. In 1986, Cambridgesoft released Chem3D, and rapidly there were many other products on the market such as RasMol, Macromodel, Chemical Design, Hypercube, CAChe Scientific all coming in that year. These computational software solutions have grown into a multi-million pounds industry with many vendors and solutions available to the chemists.

This software has enabled chemists to visualize their structures in the many creative ways. The following Figure 10 highlights a collection simple to complex inorganic structures; these inorganic structure illustrations examples have been made from a range of elements across the periodic table.

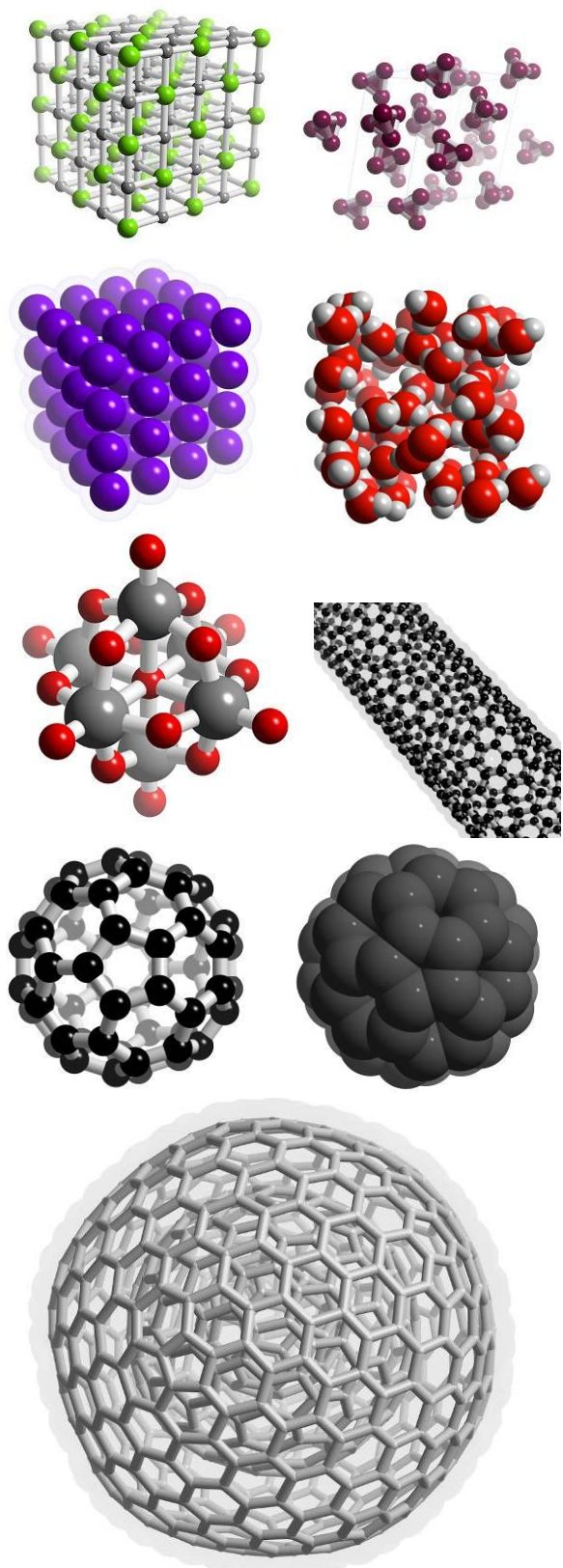

*Figure 10: 3D renders of a collection of inorganic structures*

Organic chemical structures made from predominately carbon and hydrogen can be equally beautiful as shown in the following Figure 11.

*Electronic Visualisation in Chemistry: From Alchemy to Art*
*Karl Harrison, Jonathan P. Bowen & Alice M. Bowen*

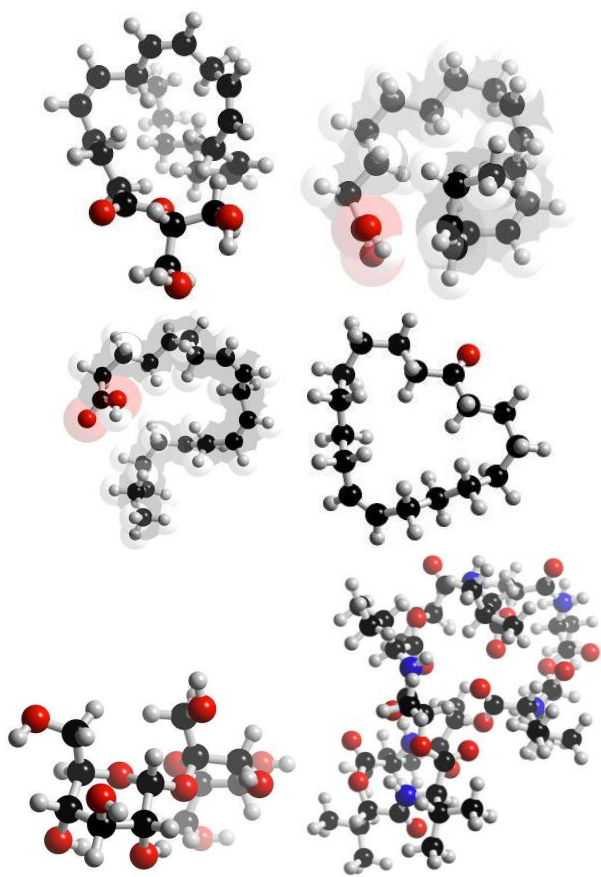

*Figure 11: 3D renders of various organic structures*

The shape and size of the actual bonds and atoms can be controlled to provide structural information. Figure 12 presents sucrose as a wire frame, a stick model, a ball and stick model, a scaled ball and stick model and finally a space-filling model, also known as a calotte model, is a type of three-dimensional molecular model where the atoms are represented by spheres whose radii are proportional to the radii of the atoms and whose centre-to-centre distances are proportional to the distances between the atomic nuclei, all in the same scale.

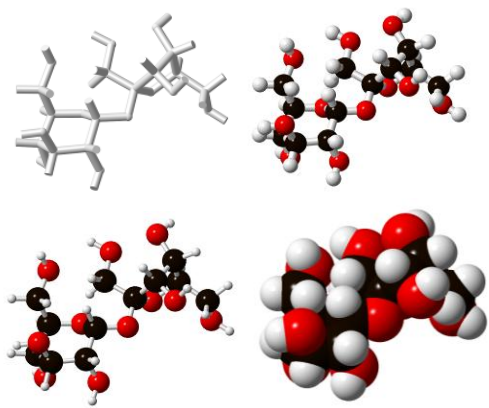

*Figure 12: Atom and bond size control in illustrations*

An interesting concept is the use of colour. Of course colour provides a useful feature in the illustrations to indicate which elements are present in the image without having to label the image. Of course the question is what colour should be used. Most people recognize a red atom to mean oxygen – but why red? What about hydrogen white and nitrogen blue? We know perhaps carbon is black. Did this colour choice come from the fact that graphite is black, but diamonds, which are solely made from carbon, are not black? How about phosphorus, it is red from the form red phosphorus or white from white phosphorus? Sulphur is easy since that element is only a pure yellow.

In the end, there turns out to be no standard convention and it is up to the illustrator to at least try to be consistent and at some point provide a label if they were to say change oxygen to purple and carbon pink. Certainly chemists are trained to look at the connectivity and recognize the elements based on what they are connected to and their patterns, but changing to "non-standard" colours is generally not helpful. The periodic table can be displayed and collection of spheres and this figure shows a standard choice of colours from Chem3D (Figure 13).

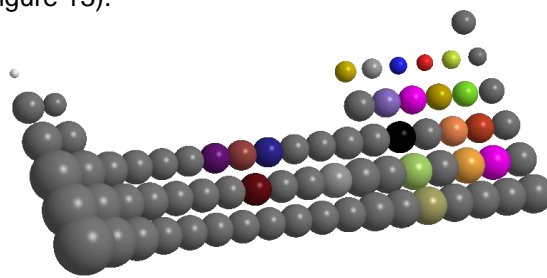

*Figure 13. Periodic table and colours of elements*

Another aspect at looking at 3D molecular structures is the concept of "secondary structure", inorganic chemists use polyhedral to provide information about repeating patterns. This can be seen in the following example, which has a wire frame, ball and stick view, atom packing view and then polyhedral view of the same structure. The polyhedral view simplifies and exemplifies the repeating network of connections in this inorganic compound (Figure 14).

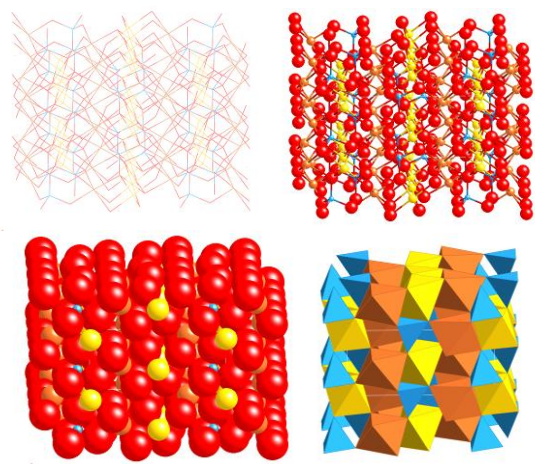

*Figure 14: Inorganic secondary structure seen in polyhedral view*

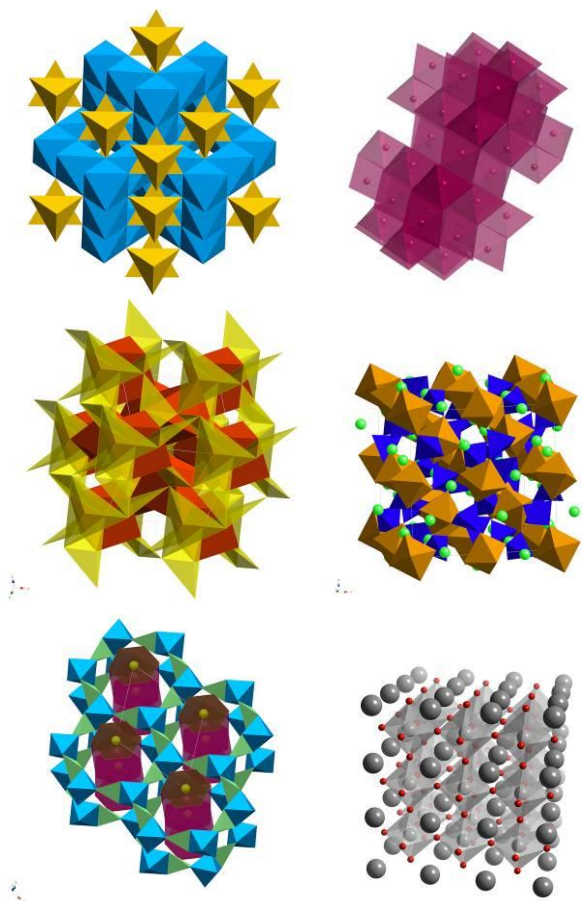

*Figure 15: Inorganic polyhedral Illustrations*

The following other inorganic polyhedral views shows the natural beauty these structure can exhibit (Figure 15).

Of course, not to be outdone, biological structures have a "secondary structure" repeating structural motif to describe how the amino acids or nucleic acid base pairs join together to form a large 3D structure. The most famous of these is the double helix of DNA (Figure 16), where the two strands of DNA bind together with complementary base pairing to form a ladder of pairs between the two spiral stands of DNA. Looking at the resulting separation of the helices in the secondary structure, one can see the major and minor groves – wider and narrower spacing – which leads to structural properties found in DNA chains and the enzymes that bind to them.

Proteins and enzymes themselves have secondary structures based on the amino acid sequences where certain sequences of amino acids give rise to sheets and others give rise to helices in the secondary structure views. It is the combination of helices and sheets which gives rise to the folded pattern which in turn provides a unique shape the protein or enzyme adopts and thus the biological chemistry that these structures perform (Figure 17).

All these colourful illustrations and 3D images that the portfolio of chemistry software packages has produced have graced in the insides of the chemistry textbooks and research publications for the last twenty years.

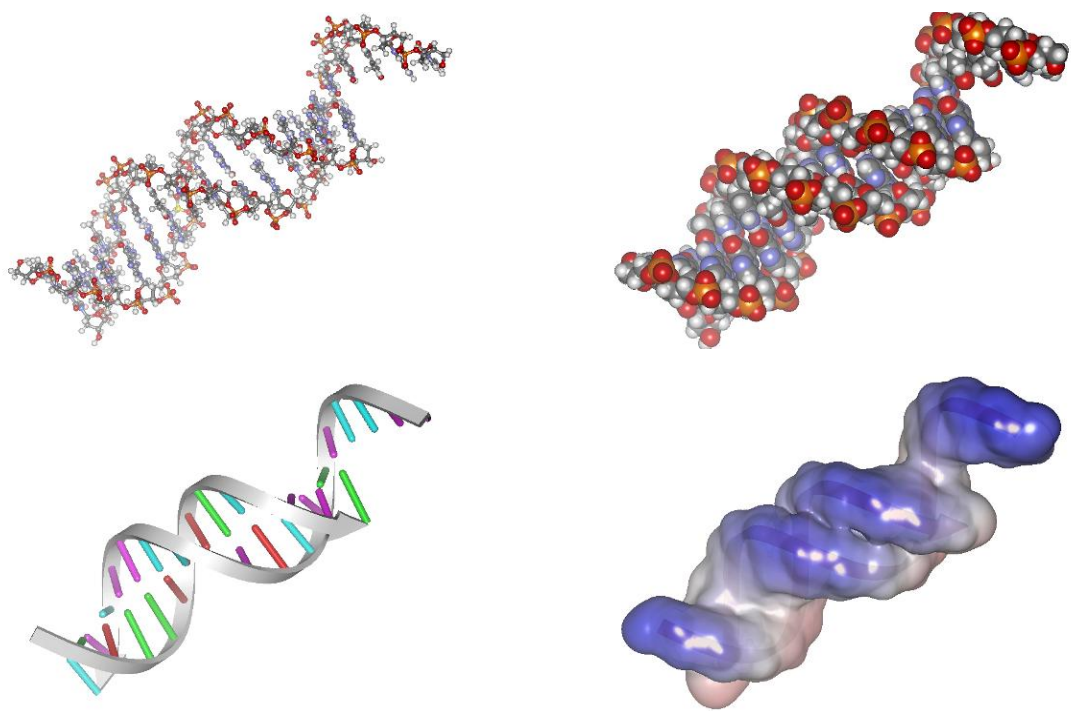

*Figure 16: Views of DNA*

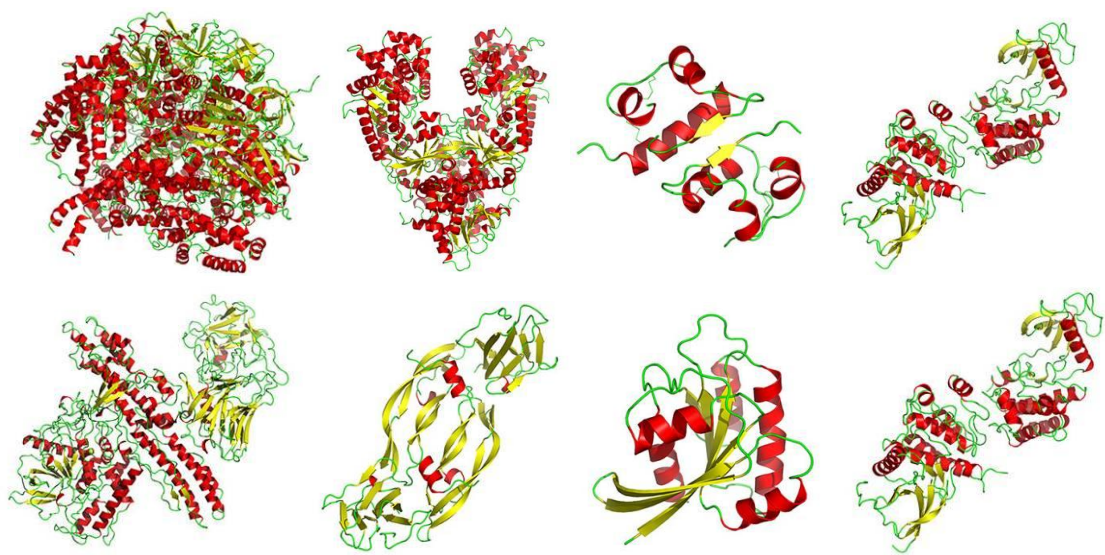

*Figure 17: Protein and enzyme secondary structures*

**JOURNAL COVERS**

More recently, journals and textbooks have been taking these colourful illustrations and using them as cover artwork (Figure 18).

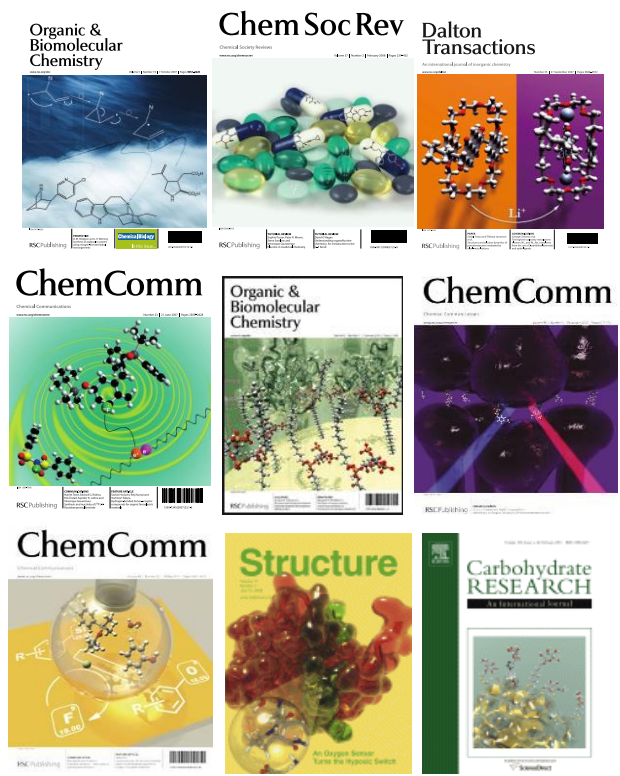

*Figure 18: Journal covers with chemistry art*

There are number of interesting comments to say regarding this process. Commercial or even learned society journal publishers will often say something like the following: "As your article was very well received by the reviewers and the editorial office, I would like to invite you to submit some eye-catching artwork for consideration for the cover of the issue in which your article will appear."

Of course, the author is flattered by the request and wants to supply an image to promote their science.

A subsequent comment from the publisher then mentions something along the following lines: "Please note that if your artwork is chosen you will be asked to make a contribution towards the production costs." This cost can run into many hundreds of pounds for the scientist to submit a cover figure and when you consider the number of journal titles a publisher has and the frequency that these titles are produced; the publisher has just then produced a new income stream of several hundreds of thousands of pounds per annum. The further comment is how many other authors has the journal written to ask for a cover, is it more than one? Or ones they think can afford to pay? Or is it really the ones that the journal wants to promote on the front because they like the science?

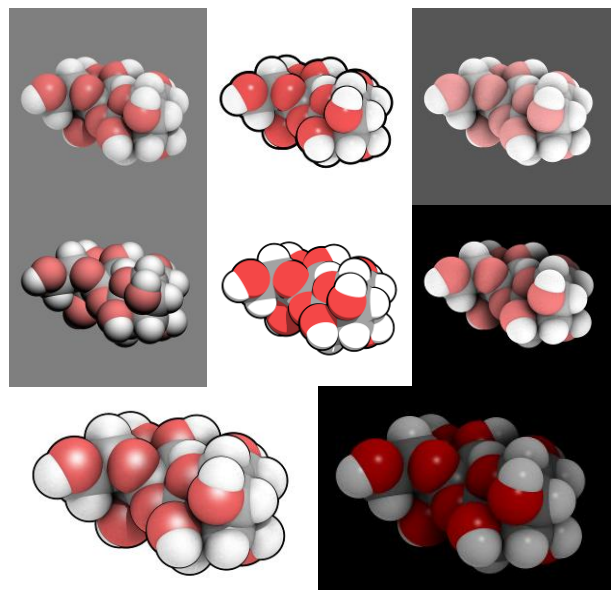

*Figure 19: 3D render lightning effects on a simple organic molecule*

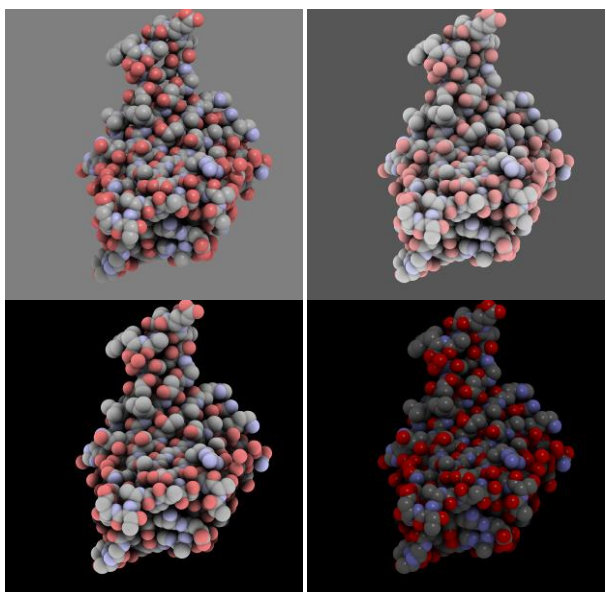

*Figure 20. 3D render lightning effects on a protein*

The competitive nature of invitations to appear on the cover has meant that artwork which now is appearing is now longer simply the output of the chemistry software products and examples shown above, but generally are now produced using high-level technical skills with tools like Adobe Illustrator, or 3D production software solutions such as Autodesk Maya, PovRay, 3D Studio Max, CINEMA 4D Studio (Figure 19 and 20).

Of course the expectation that a research chemist or textbook author could have the skills to use these software solutions outside their normal research computer tools is low, so the scientists have turned full circle now and are collaborating with experts in using electronic visualisation tools, such as 3D artists. One of this paper's authors, Karl Harrison, has providing such a role for members of the Department of Chemistry at the University of Oxford (Figure 18 & 21).

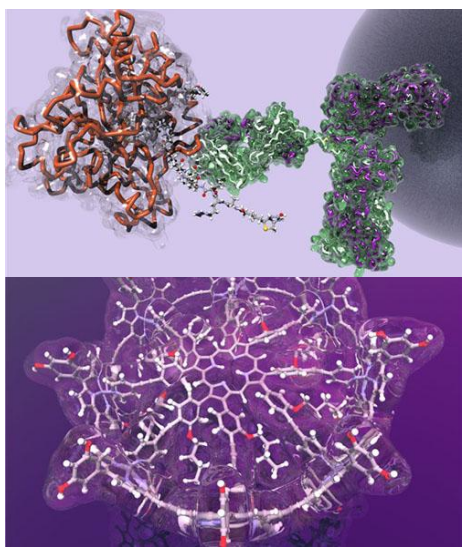

*Figure 21: 3D renders of chemical structures*

The following figures 22–28 illustrate a typical discussion and design from initial science illustration in a research paper to cover image. A research scientist at Oxford had been invited to submit a cover illustration based on their novel science – which was the single step biotinylation of 5hmC in 100mer ssDNA with N-biotinyl L-cysteine and this was then enrichment and allowed isolation of biotinylated DNA using an enzyme modified nanoparticle bead Dynabeads® M-280 streptavidin. The scientist provided an initial concept for the cover art Figure 22.

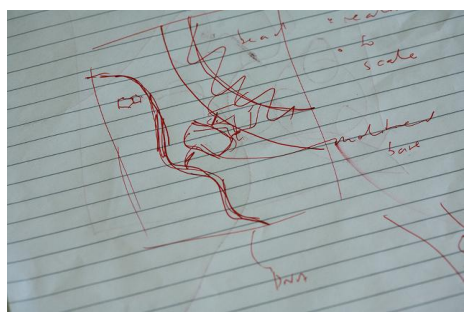

*Figure 22: Initial concept sketch*

An electronic visualisation and resulting cover art therefore needed a number of components, a 3D model of the 100 unit single strand of DNA (Figure 23), modified this with the chemical addition (Figure 24), a 3D model of the protein streptavidin (Figure 25) and then an understanding of the coverage of the streptavidin protein on the nanoparticle bead.

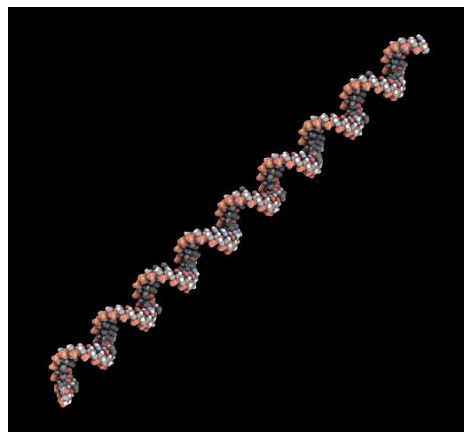

*Figure 23: Component 1 for cover art*

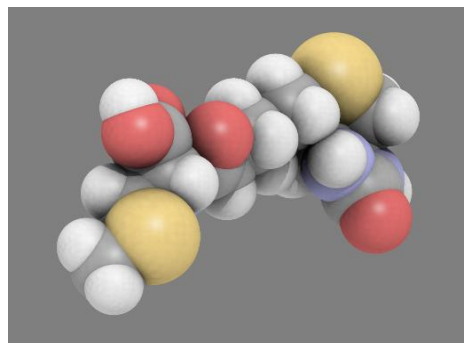

*Figure 24: Component 2 for cover art*

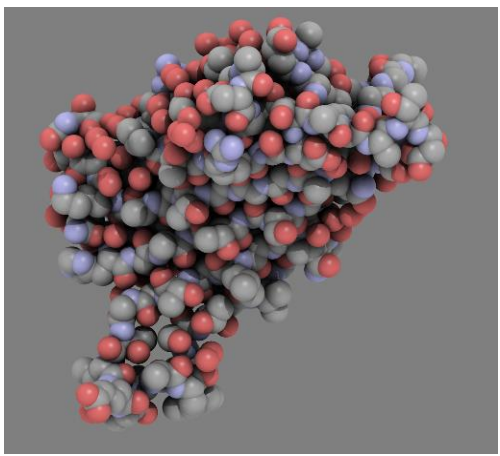

*Figure 25:* Component 3 for cover art

Combining all these elements together in a 3D ray-tracing program provided an initial draft shown in Figure 26 and then this was modified to build in 3D layers to produce the final illustration for the cover, which was accepted for by the journal editors Figure 27.

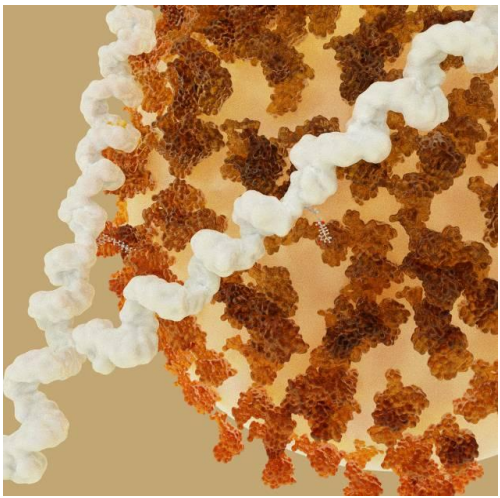

*Figure 26.* Initial draft design

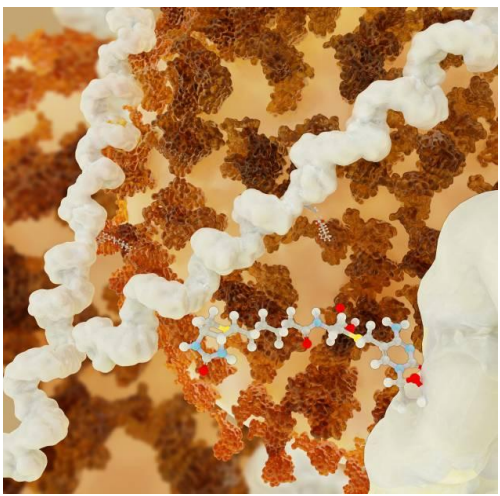

*Figure 27.* Cover design

## CONCLUSION

The combinations of facilities for handling 3D meshes, materials, instancing, depth of field, reflection mapping, High-Dynamic Range Imaging (HDRI) and Image Based Lighting (IBL), which are available in 3D rendering software solutions, can provide highly abstract or very photorealistic illustrations. This has led to an exciting time where scientists are working with artists using electronic visualisation to see their chemistry in new ways and also to show the public the beauty in the science they are undertaking (Phipps 2008).

There is a wide variety of software used for the visualisation of molecules. These can be very complex, such as enzymes, proteins, etc., and even illustrations for scientific use can have artistic aspects for a non-specialist viewer; e.g., see Bowen (2013), which inspired the origins of this paper. A full survey of chemistry visualisation software cannot be included here for space limitation reasons. However, for those interested in the possibility of using such software for artistic purposes, Wikipedia provides a list of molecular graphics software systems (Wikipedia 2013).

In summary, this paper illustrates the aesthetic possibilities of visualisation in chemistry (Hoffmann 2003), especially at the molecular level, the historical development of illustrating molecules, and the current state of the art in this area. It is hoped that this may inspire artists to undertake future interdisciplinary collaboration with chemists from a more purely artistic standpoint and to utilise the wide selection of available visualisation software for artistic purposes.